\documentclass[twocolumn,showpacs,prl]{revtex4}
\usepackage{graphicx,amsmath}
\newcommand{\ket}[1]{|#1\rangle}
\newcommand{\bra}[1]{\langle #1|}
\newcommand{\figref}[1]{Fig.~\ref{#1}}
\begin{document}
\title{Cavity QED with separate photon storage and qubit readout modes}
\author{P.~J.~Leek}
\altaffiliation{These authors contributed equally to this work.}
\author{M.~Baur}
\altaffiliation{These authors contributed equally to this work.}
\author{J.~M.~Fink}
\author{R.~Bianchetti}
\author{L.~Steffen}
\author{S.~Filipp}
\author{A.~Wallraff}
\email{andreas.wallraff@phys.ethz.ch}
\affiliation{Department of Physics, ETH Zurich, CH-8093, Zurich, Switzerland.}

\pacs{03.67.Lx, 03.67.Bg, 42.50.Pq, 85.25.-j}

\date{\today}
\begin{abstract}

We present the realization of a cavity quantum electrodynamics setup in which photons of strongly different lifetimes are engineered in different harmonic modes of the same cavity. We achieve this in a superconducting transmission line resonator with superconducting qubits coupled to the different modes. One cavity mode is strongly coupled to a detection line for qubit state readout, while a second long lifetime mode is used for photon storage and coherent quantum operations. We demonstrate sideband based measurement of photon coherence, generation of $n$ photon Fock states and the scaling of the sideband Rabi frequency with $\sqrt{n}$ using a scheme that may be extended to realize sideband based two-qubit logic gates.
\end{abstract}
\maketitle

\begin{figure}[b]
\includegraphics[width=1.0 \columnwidth]{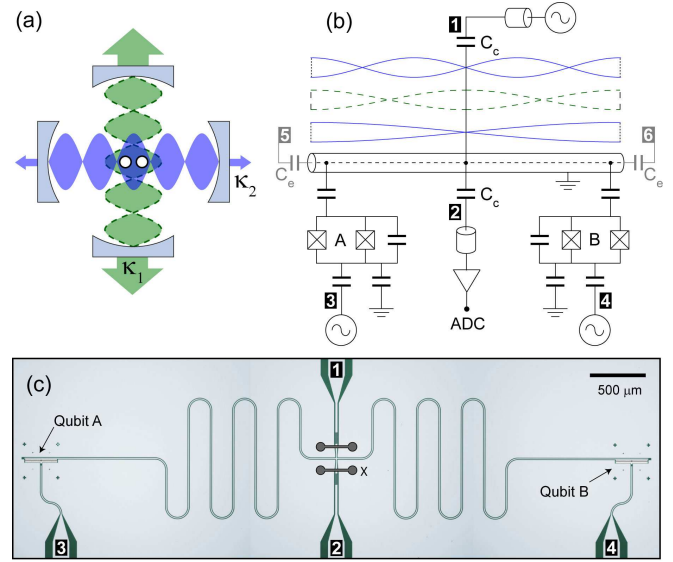}
\caption{(Color online) (a) Crossed optical Fabry-Per\^ot cavities geometry with two atoms for experiments with two different photon lifetimes. (b) Circuit diagram of center coupled transmission line resonator with qubits (A, B) coupled at each end. The resonator is driven through port 1, transmission is measured at port 2. Qubits can be directly driven through ports 3 and 4. Resonator end ports 5 and 6 were included only in a test device without qubits. Also depicted are the field amplitude of the first three resonator modes: odd modes (blue, solid), even modes (green, dashed). (c) Optical image (colorized) of the device, with two transmon qubits. Grounding bond wires are shown schematically at position X.} \label{one}
\end{figure}

In cavity quantum electrodynamics (QED) \cite{Walther2006,Haroche2007,Ye2008} the interaction between matter and light confined in a cavity is studied. The lifetime of a photon in a cavity can be engineered by changing the transparency of the cavity mirrors. One way to carry out experiments with photons of different lifetimes would be to use two different cavities, for example in a crossed Fabry-Per\^ot geometry (see \figref{one}). Here we demonstrate a method of doing such experiments by instead controlling the quality factors of different harmonic modes of a single cavity, in an electrical implementation of cavity QED known as circuit QED \cite{Wallraff2004,Schoelkopf2008}. In circuit QED very large coupling strengths can be achieved between macroscopic superconducting qubits and a strongly confined microwave field in an on-chip transmission line resonator \cite{Wallraff2004,Schoelkopf2008}. This has formed the basis of remarkable recent progress in quantum information processing (QIP) \cite{Dicarlo2009,Ansmann2009} and quantum optics \cite{Houck2007,Fink2008,Schuster2007,Hofheinz2008,Fink2009,Hofheinz2009,Bishop2009} with superconducting circuits. The quasi one-dimensional cavity in this system allows for a novel geometry to be used here that is unavailable in conventional optical or microwave frequency cavity QED.

The lifetime of resonator photons in circuit QED is controllable via capacitive coupling of input and output transmission lines at the resonator ends \cite{Goeppl2008}. Here we instead couple input/output lines at the resonator center. A circuit schematic and optical image of the device with two embedded superconducting transmon qubits \cite{Koch2007,Schreier2008} are shown in \figref{one} (b and c). The first harmonic and other even-symmetry modes have an electric field antinode at the center, and are hence strongly coupled to the center ports 1 and 2. Conversely, the fundamental and higher odd-symmetry harmonic modes have an electric field node at the center of the resonator and couple only weakly to the center ports \cite{BondNote}.  Choice of the capacitance between the resonator center and the external lines allows one to define an external quality factor (Q), at the even resonator modes, leaving the Q factor of the odd modes almost unaffected, and limited only by internal losses. Embedding superconducting qubits into such a device allows us to perform cavity QED experiments with two radically different photon lifetimes in the same resonator.

A transmission measurement through a strongly coupled resonator mode can be used to perform a quantum non-demolition measurement of a qubit detuned from the resonator \cite{Wallraff2005}. This technique has been extended to perform simultaneous joint readout and complete tomography of two qubits \cite{Filipp2009} and has also been used for readout in the realization of two-qubit quantum algorithms \cite{Dicarlo2009}. For this dispersive measurement to be efficient the cavity lifetime should be dominated by coupling into the detector port rather than by internal loss and much shorter than the qubit lifetimes to gain information about the qubit states before relaxation occurs \cite{Gambetta2008}.

This requirement can often be counter to a desire for maximized photon lifetimes for cavity QED and QIP experiments. An example of this trade-off occurs when one wishes to use photons to remotely couple two qubits, by for example tuning qubits into resonance with a cavity \cite{Hofheinz2009,Ansmann2009}, or via resonant sideband transitions \cite{Leek2009}. The fidelity of entangled qubit states generated using sideband transitions in our own experiments has so far been limited by a photon lifetime chosen to also allow efficient qubit measurement \cite{Leek2009}. The problem can be solved if additional circuitry separate from the coupling resonator is used for measurement, either in the form of individual qubit readouts \cite{Ansmann2009,Mallet2009}, or a separate joint readout resonator. However, utilizing the multiple harmonic modes of a single resonator and engineering their Q factors represents a viable solution which enables the coupling of two or more qubits to both modes simultaneously with little increase of circuit complexity.

\begin{figure}[t]
\includegraphics[width=1.0 \columnwidth]{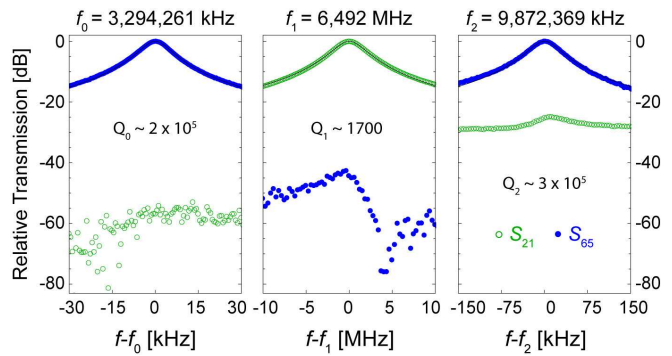}
\caption{(Color online) Normalized transmission spectrum for a 4-port center and end-coupled resonator (device I). Transmission between the end ports $S_{65}$ (blue, closed circles) and the center ports $S_{21}$ (green, open circles) at the fundamental (a), 1st (b) and 2nd harmonic (c) are shown.} \label{two}
\end{figure}

For initial assessment of the center coupling concept, a device (I) was fabricated with weakly capacitively coupled end ports 5 and 6 (see \figref{one}b) but without qubits. The capacitances were chosen small enough ($C_{\rm{e}}\simeq1~\rm{fF}$) to have little effect on the quality factor of the first three harmonic modes, while the center ports were strongly coupled ($C_{\rm{c}}=15~\rm{fF}$). Measurements of these first 3 harmonics are shown in \figref{two}. The data is normalized to a resonant transmission of $0~\rm{dB}$ in each case. The center coupled transmission ($S_{21}$, green) shows a resonance at $f_1\approx6.49~\rm{GHz}$ with quality factor $Q_1\approx1700$ (\figref{two}b) consistent with the chosen value of $C_{\rm{c}}$, but low transmission around $f_0$ and $f_2$. The end-coupled transmission ($S_{65}$, blue) displays resonances at $f_0\approx3.29~\rm{GHz}$ (a), and $f_2\approx9.87~\rm{GHz}$ (c) with much higher measured quality factors $Q_0\approx3\times10^5$ and $Q_2\approx2\times10^5$ respectively \cite{Qnote}, and low transmission at $f_1$ (b). These Q factors are on the same order as those measured for weakly end-coupled two-port resonators \cite{Oconnell2008,Goeppl2008}, showing that the fundamental and 2nd harmonic in the 4-port device are negligibly coupled to the center ports. Hence the device performs as desired, with the first harmonic strongly coupled to the outside world for qubit readout, and the odd modes well isolated for optimal photon storage times.

A second device (II) fabricated with a transmon qubit at each end of the resonator (see \figref{one}c) was used for all further experiments. This device has higher mode frequencies than device A, and no end ports (5 and 6). The qubits both have a charging energy $E_{\rm{C}}/h\approx 305~\rm{MHz}$, a maximum Josephson energy $E_{\rm{Jmax}}/h\approx 150~\rm{GHz}$, and flux tunable transition frequencies to a maximum of around $19~\rm{GHz}$. Direct microwave drive lines (ports 3 and 4) allow selective driving of the individual qubits.

The resonant coupling strengths $g_1/{2\pi}=119\pm1~\rm{MHz}$ of the qubits to the 1st harmonic at $\omega_1/{2\pi}=7.01~\rm{GHz}$ were extracted from a standard resonator transmission $S_{21}$ measurement of the vacuum Rabi mode splitting \cite{Wallraff2004,Fink2009}, using a locally coupled flux bias coil to tune the qubit frequency. The coupling strength $g_{\rm{2}}/{2\pi}=183\pm5~\rm{MHz}$ to the 2nd harmonic at $\omega_2/{2\pi}=10.74~\rm{GHz}$ was instead obtained from a spectroscopic measurement of the qubit transition frequency \cite{Schuster2005}. In this dispersive measurement the phase shift of a fixed frequency tone at the first harmonic of the resonator was monitored as the transition frequency of each qubit was swept across the 2nd harmonic resonance frequency of the resonator.

\begin{figure}[b]
\includegraphics[width=0.96 \columnwidth]{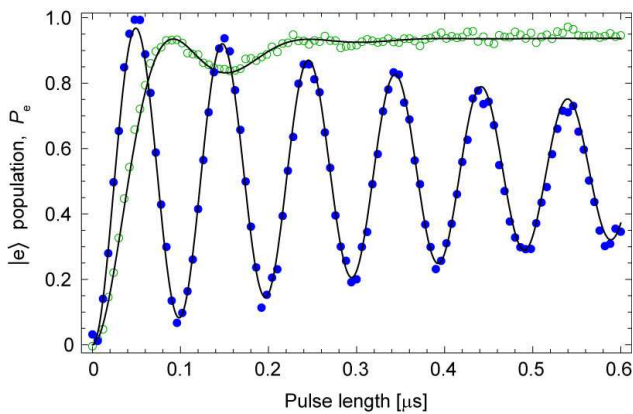}
\caption{(Color online) Rabi oscillations on the blue sideband with the low Q 1st harmonic (green, open circles), and with the high Q 2nd harmonic (blue, closed circles). Master equation simulations with the photon lifetime and sideband drive rate as fit parameters are shown as solid lines.} \label{three}
\end{figure}

To demonstrate the functionality of the device, we couple the qubits to the different cavity modes using sideband transitions \cite{Wallraff2007}. We carry out experiments in the dispersive regime in which the qubits are far detuned from both modes. The qubit population is in all cases extracted from a dispersive measurement at the 1st harmonic of the resonator \cite{Bianchetti2009}. We first tune one qubit (A) to a transition frequency of $5.49~\rm{GHz}$, at a detuning $\Delta/{2\pi}=1.52~\rm{GHz}$ below the 1st harmonic. We then drive the two photon blue sideband with a microwave field applied to port 3 and observe Rabi oscillations between the states $\ket{g,0}$ and $\ket{e,1}$, where $g$ and $e$ are the ground and first excited states of the qubit respectively, and $0,1$ refer to the photon number in the resonator mode (see \figref{three}) \cite{Leek2009}. The rapidly decaying Rabi oscillations fit well to a master equation simulation, with a short photon lifetime of $T_1^{\kappa1}=39~\rm{ns}$ that is consistent with the linewidth of the resonator spectrum. The sideband drive rate is extracted as $\Omega^{\kappa1}/{2\pi}=6.9~\rm{MHz}$. Since $T_1^{\kappa1}$ is much shorter than the qubit lifetime $T_1= 1.0~\rm{\mu s}$, the qubit excited state population saturates at long times at $P_e \approx 0.93$, an effect that efficiently creates population inversion \cite{Leek2009}. An equivalent experiment is then carried out with the qubit at a transition frequency of $9.22~\rm{GHz}$, at the same detuning $\Delta$ but this time below the 2nd harmonic. The drive rate in this case was $\Omega^{\kappa2}/{2\pi}=9.8~\rm{MHz}$ and the qubit lifetime $T_1= 730~\rm{ns}$. High contrast sideband Rabi oscillations are observed in this case due to the much longer photon lifetime $T_1^{\kappa2}=1.6~\rm{\mu s}\gg T_1^{\kappa1}$ in this cavity mode.

\begin{figure}[t]
\includegraphics[width=0.88 \columnwidth]{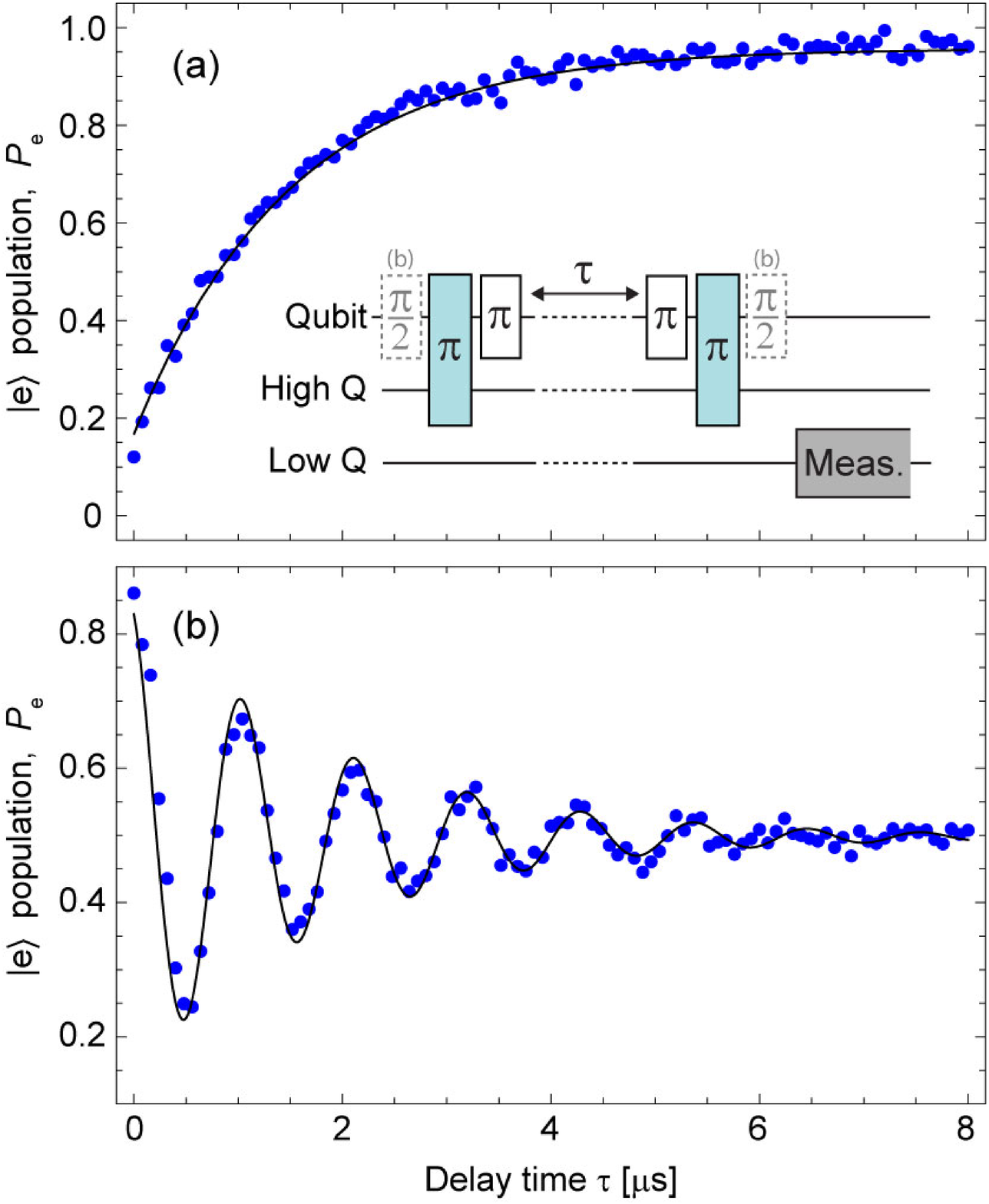}
\caption{(Color online) (a) Measurement of the lifetime $T_1^{\kappa2}$ of a single photon in the high Q mode. Inset: Pulse schematic for the experiment with sideband pulses shaded in blue. (b) Results of a Ramsey experiment to measure the photon dephasing time $T_2^{*\kappa2}$. The pulse sequence is identical to that in (a), with the addition of two $\pi$/2 pulses as shown in the inset of (a). Master equation simulations are shown as solid lines.} \label{four}
\end{figure}

Using a sequence of blue sideband and direct qubit pulses, we now demonstrate a photon storage experiment with the long lifetime mode. The system is first excited to the state $\ket{e,1}$ using a blue sideband $\pi$-pulse. A direct $\pi$-pulse on the $g-e$ transition returns the qubit to the ground state and leaves the resonator in a single photon Fock state $\ket{g,1}$. After a storage time $\tau$, the two pulses are repeated in reverse order. For perfect photon storage the qubit population should return to state $\ket{g}$. However, when the photon is lost the qubit ends up in state $\ket{e}$. The measured result is shown in \figref{four}(a), along with a master equation simulation fitted to the data, with a photon lifetime of $T_1^{\kappa2}=1.45~\rm{\mu s}$, close to the value found from the sideband Rabi data in \figref{three}. This corresponds to a quality factor of $Q=97000$. We also measure the photon dephasing time using a Ramsey type pulse sequence (\figref{four}(b)), finding $T_2^{*\kappa2}=1.9~\rm{\mu s}$. The fact that $T_2^{*\kappa2}<2T_1^{\kappa2}$ indicates the presence of some fluctuation of the resonator frequency, which may be partially accounted for by the dispersive coupling to the qubit, which at the chosen transition frequency has a separately measured dephasing time of $T_2^*\simeq 250~\rm{ns}$. The excellent coherence properties of such a long lived cavity mode could enable its use as a quantum memory, collectively accessible to multiple qubits.

\begin{figure}[t]
\includegraphics[width=1.0 \columnwidth]{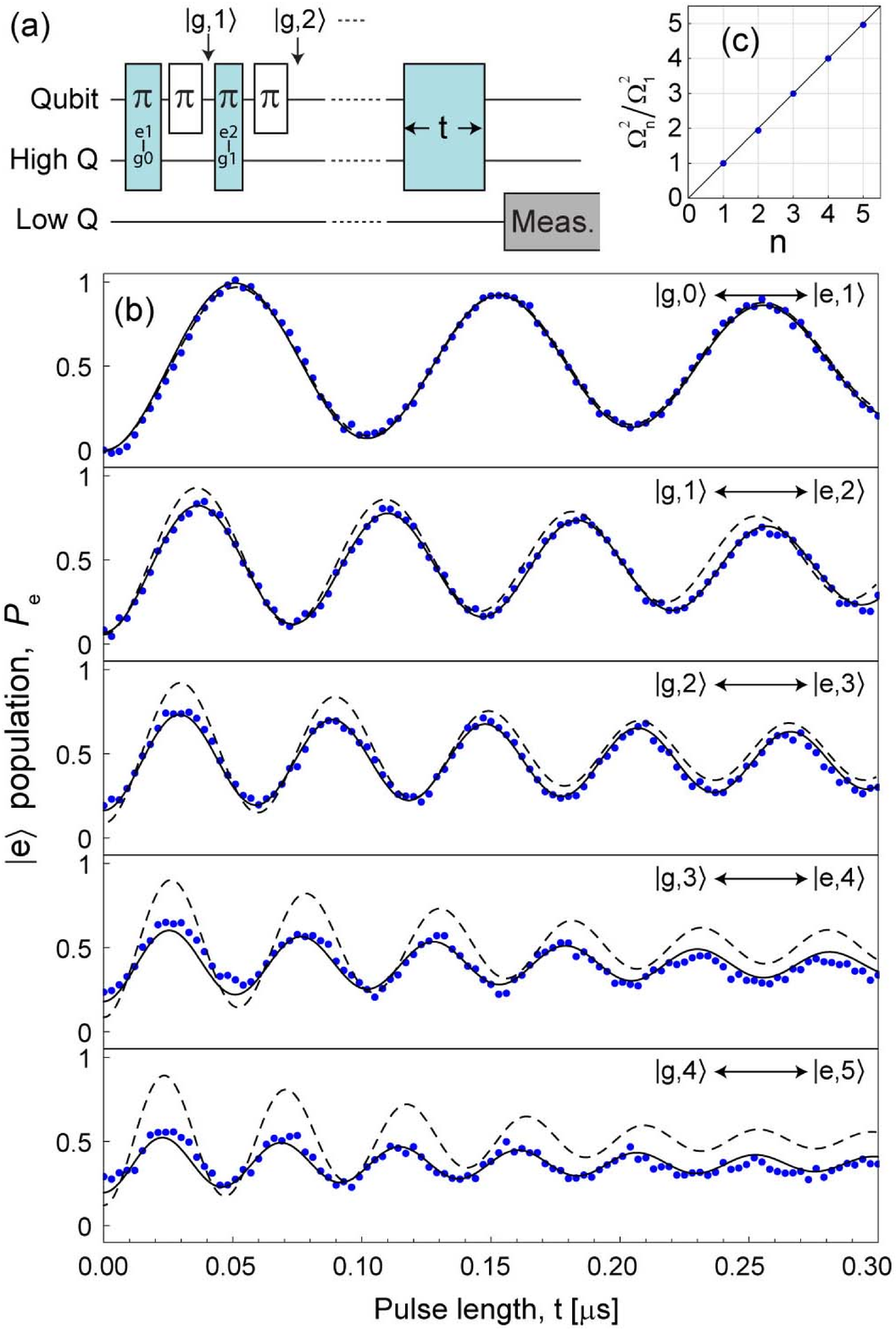}
\caption{(Color online) (a) Pulse sequence for generation of Fock states in the high Q mode, and driving of sideband Rabi oscillations $\ket{g,n-1}\longleftrightarrow\ket{e,n}$. (b) Measurements for the sequence depicted in (a), for $n=1,2,3,4,5$ (blue points). Fits to a cosine with an exponential decay envelope are shown as solid lines, while master equation simulations are shown as dashed lines. (c) Plot of $\Omega_n^2/\Omega^2_1$ vs $n$ (blue points), where $\Omega_n$ is the Rabi frequency extracted from the fits shown in (b). Expected dependence $\Omega_n^2/\Omega^2_1=n$ is shown as a black line.} \label{five}
\end{figure}

The high Q of the 2nd harmonic also allows us to carry out more complex sideband pulse sequences to generate Fock states $\ket{n}$ of the long-lived microwave field, and to observe Rabi oscillations on the blue sideband between the states $\ket{g,n-1}$ and $\ket{e,n}$ up to $n=5$ (see \figref{five}). The state $\ket{g,n}$ is generated by applying a sideband $\pi$-pulse followed by a direct qubit $\pi$-pulse and repeating $n$ times (see \figref{five}a). In \figref{five}b we show the results of sideband Rabi oscillation experiments starting from the experimentally generated Fock states with $n=1,2,3,4,5$. A fit to $P_{\rm{e}}(t)=A-Be^{-t/\tau}\cos(\Omega_n t)$ yields Rabi frequencies $\Omega_n$ that are in very close agreement with the expected $\sqrt{n}$ scaling of the coupling strength (see \figref{five}c) \cite{Fink2008,Hofheinz2008}. Master equation simulations are also shown in \figref{five}b, agreeing qualitatively with the measured data, but deviating for the longer pulse sequences. This is likely due to build up of errors due to off-resonant driving of other transitions, and will be important to characterize and correct before taking operation complexity further in future experiments. A good understanding of these higher sideband transitions should allow their use as the basis of a universal two-qubit gate \cite{SchmidtKaler2003}.

An attractive feature of the two-mode device is the possibility to carry out joint dispersive measurement and photon based operations on two qubits coupled to both modes simultaneously, a task that would be more difficult with multiple cavities. In order to demonstrate this, we have generated Bell states between two qubits in this device using a sideband scheme \cite{Leek2009}, and characterized the resulting states using state tomography \cite{Filipp2009}. Measured states $\rho$ were found to have fidelities $\mathcal{F}\equiv(\bra{\psi}{\rho}\ket{\psi})^{1/2}$ of $\mathcal{F}=87\%$ and $86\%$ and concurrence \cite{Horodecki2008} $\mathcal{C}=0.52$ and $0.51$ with respect to the ideal states $\ket{\psi}=\ket{\Psi_+}=(\ket{ge}+\ket{eg})/\sqrt{2}$ and $\ket{\Phi_+}=(\ket{gg}+\ket{ee})/\sqrt{2}$. The fidelity of the prepared states is now not limited by photon lifetime but rather by the the length of the entangling sideband pulses relative to the qubit dephasing times. Shorter side band pulses would require higher drive rates, that are observed to lead to off-resonant driving of direct qubit transitions reducing the fidelity of the prepared states. This aspect will likely be improved in future experiments using optimal control techniques.

In conclusion, we have realized a microwave frequency on-chip resonator for cavity QED experiments with photon lifetimes in adjacent harmonic modes differing by a large factor. We have demonstrated photon storage, Fock state and Bell state generation with long lived photons of one mode, while carrying out efficient dispersive qubit readout using short lived photons in a second mode. Such mode lifetime engineering enables increased flexibility in the use of harmonic modes for qubit-qubit coupling in superconducting circuits, and in the control of decoherence in such systems. The concept could thus be useful in the design of future quantum information processors.

We thank A. Blais for valuable discussions and comments on the manuscript. We acknowledge the group of M. Siegel at the University of Karlsruhe for the preparation of Niobium films. This work was supported financially by the Swiss National Science Foundation, by the EC via the EuroSQIP project and by ETH Zurich.

\vspace*{-5mm}

\bibliographystyle{apsrev}

\begin{thebibliography}{23}
\bibitem{BondNote}
To avoid disruption of the odd mode fields, the ground planes must be well connected across the center ports, achieved here with bond wires.
\end{thebibliography}

\end{document}